\documentstyle[12pt,emulateapj,psfig]{article}

\newcommand{\gtilde}
 {~ \raisebox{-1ex}{$\stackrel{\textstyle >}{\sim}$} ~}
\newcommand{\ltilde}
 {~ \raisebox{-1ex}{$\stackrel{\textstyle <}{\sim}$} ~}

\begin{document}

\title{Galaxy Formation by Galactic Magnetic Fields}

\author{Tomonori Totani}
\affil{ National Astronomical Observatory, Mitaka, Tokyo 181-8588,
Japan \\
E-mail: totani@th.nao.ac.jp
}

\submitted{
To Appear in ApJ Letters; 
Received March 2; Accepted March 26}

\renewcommand{\baselinestretch}{1.2}

%\small
\footnotesize
%\scriptsize

\begin{abstract}
Galaxies exhibit a sequence of various morphological types, i.e., the Hubble
sequence, and they are basically
composed of spheroidal components (elliptical galaxies and bulges
in spiral galaxies) and disks. %22
It is known that spheroidal components are found
only in relatively massive galaxies with $M=10^{10-12}M_\odot$, and
all stellar populations in them are very old,
but there is no clear explanation for these facts. %27
Here we present a speculative scenario for the origin of the
Hubble sequence, in which magnetic fields
ubiquitously seen in galaxies have played a crucial role. %26
We first start from a strange observational fact that
magnetic field strengths observed in spiral galaxies
sharply concentrate at a few microgauss, for a wide range of galaxy
luminosity and types. We then argue that
this fact and the observed correlation between star formation activity and
magnetic field strength in spiral galaxies suggest that
spheroidal galaxies have formed by starbursts induced by strong 
magnetic fields. Then we 
show that this idea naturally leads to the formation of 
spheroidal systems only in massive and high-redshift objects in
hierarchically clustering universe,
giving a simple explanation for various observations. %49
\end{abstract}


\section{Introduction}
It is generally believed that
the hierarchical structure today seen in the universe is
generated by gravitational instability of cold dark matter (CDM) whose
density is about 10 times higher than that of baryonic matter,
and galaxies form in virialized dark matter haloes
by cooling of baryonic gas and subsequent star formation.
The characteristic mass range of galaxies is about $10^{8-12}M_\odot$,
which can be understood as a mass range in which baryonic gas can cool
sufficiently within a dark matter halo to form a galaxy (e.g., Blumenthal 
et al. 1984 for a review). However,
spheroidal components reside only in relatively massive galaxies with
$M = 10^{10-12} M_\odot$ and less massive galaxies are mostly irregular
types or late-type spiral galaxies (Burstein et al. 1997; Hunter 1997). 
[There is a population called
as dwarf spheroidal galaxies, but they are generally considered to be
different from giant elliptical galaxies and bulges. They are 
more similar to dwarf irregulars (Bender et al. 1992; Mao \& Mo 1998).] 
Elliptical galaxies and bulges
appear to form a very uniform, old stellar population with very little scatter
in metallicity and star formation history, and they are generally
considered to have formed by intensive starbursts at high redshifts
(e.g., Renzini 1999). The trigger mechanism for starbursts is unknown, and 
it is often assumed to be mergers of disk galaxies when galaxy
formation is modeled in the context of hierarchical structure formation
in the CDM universe (Kauffman, White, \& Guiderdoni 1993; 
Baugh, Cole, \& Frenk 1996). However, there is no evidence for 
spheroidal galaxy formation through mergers, and the merger hypothesis
does not give a clear explanation for the fact that all spheroidal
galaxies are very old and only in massive galaxies. These trends appear
to run opposite to the expectation of structure formation in the CDM
universe, in which smaller objects form earlier and then massive
objects later. This is currently considered as a major challenge for 
galaxy formation in the CDM universe (e.g., Renzini 1999), and it is
interesting to seek for a mechanism which triggers starbursts only
in massive and high-redshift objects. In this letter we show that a possible
strong dependence of star formation activity on magnetic field
strengths, which is suggested by observations of magnetic fields in
local galaxies, gives a candidate for such a mechanism.

\section{Galactic Magnetic Fields and Star Formation Activity}
Interstellar magnetic fields in galaxies
are strong enough to significantly affect
interstellar gas dynamics on both global scales characteristic of
galactic structure and small scales characteristic of star formation
(see, e.g., Vall\'ee 1997;
Zweibel \& Heiles 1997 for a review). Fitt and Alexander (1993; hereafter FA) 
derived strengths of volume-averaged magnetic fields
for 146 spiral galaxies, which is a complete sample 
including various types from Sa to irregular galaxies, and they found a 
strange result: the dispersion in strength is surprisingly narrow
within a factor of less than 
2 with an average strength of 3 $\mu$G, in spite of
a wide range of absolute radio luminosity of galaxies (almost 4 orders
of magnitude). Why is magnetic field strength so uniform
at $\sim 3 \mu$G for various types of galaxies with such a wide range of
luminosity? Clearly there should be a physical reason for this strange fact. 

Vall\'ee (1994) investigated the relation between the magnetic
field strength and star formation activity by using 48 galaxies out of
the FA sample, and found 
that there is a correlation between magnetic field strength ($B$) 
and star formation efficiency (SFE) 
as B $\propto ({\rm SFE})^i$ with $i=0.13 \pm 0.04$. 
[The same relation was found also
for star formation rate (SFR) and $B$.]
From this result,
one may conclude that star formation activity does not strongly affect the
strength of magnetic fields. However,
it is also likely that star formation rate {\it is} 
affected by magnetic fields. If we take the result of Vall\'ee in this
context, the observed correlation suggests that star formation
activity depends quite sensitively on $B$ as SFE $\propto B^q$ 
with $q = 1/i \sim 7.7 \pm 2.6$. 
Here we point out that, if this is the case,
the surprisingly narrow dispersion in 
galactic magnetic fields is naturally explained 
by an observational selection effect.
If the strength of magnetic field in an object 
is significantly smaller than a few $\mu$G, the
star formation activity becomes much lower than in typical spiral
galaxies, and such object cannot produce enough stars to be observed as
a galaxy.
On the other hand, if magnetic fields are stronger than a few $\mu$G,
such objects would experience strong starburst and all interstellar 
gas will be converted into stars. If the star formation time scale is
shorter than the dynamical time scale for disk formation,
they will become present-day spheroidal systems.
Magnetic fields in
elliptical galaxies are difficult to measure since there are no
relativistic electrons to illuminate the magnetic fields by synchrotron
radiation. Then the reason why the observed dispersion of magnetic fields
is quite small can be understood by an observational selection effect; 
we cannot observe or
measure magnetic field strength when the strength is significantly different
from the typically observed value of $\sim 3 \mu$G.

Therefore the observed correlation between $B$ and
SFE, as well as the surprisingly narrow dispersion of
measured values of magnetic field strength, 
well suggests a strong dependence of 
star formation activity on magnetic field strength.
This idea is further supported by an amorphous spiral galaxy M82, 
which has abnormally large
magnetic field of $\sim 10 \mu$G in the sample of FA.
M82 is actually known as an archetypal starburst galaxy,
and the strong field strength suggested by FA is caused by the very
strong field ($\sim 50 \mu$G)
in the nuclear region of this galaxy, where intense star formation is
underway (Klein, Wielebinski, \& Morsi 1988; Vall\'ee 1995a).
Unfortunately our knowledge for physics of star
formation is poor and it is not clear from a
theoretical point of view why SFE depends so strongly on $B$, but
it is not unreasonable. In fact, at least one effect is known by which
magnetic fields help star formation: the magnetic braking (e.g., 
McKee et al. 1993 for a review). 
In order for a protostellar gas cloud to collapse into stars, a significant
amount of initial angular momentum must be transported outward.
Magnetic fields play an important role for this angular momentum loss
by Alfv\'en waves launched into ambient medium. Strongly magnetized
objects with large scale high gas density would be followed by 
magnetic braking of individual regions with angular momentum losses,
ending in outgoing Alfv\'en waves, gas collapse, and star formation on
a large scale.

It should be noted that the above hypothesis does not contradict
with the well-known Schmidt law of star formation. Vall\'ee (1995b, 1997)
found a relation between magnetic field strengths and interstellar 
gas density on a scale of a whole galaxy as $B \propto n^k$ with $k =
0.17 \pm 0.03$. By using this relation and $B$-SFR relation,
we find SFR $\propto n^{k/i}$ with $k/i = 1.3 \pm 0.1$. 
This relation agrees well (Vall\'ee 1997)
with some direct estimates of the relation
between SFR and $n$ (e.g., Kennicutt 1989; Shore \& Ferrini 1995; 
Niklas \& Beck 1997). At the same time, these relations suggest that
the thermal energy of interstellar matter is not in equipartition
with the magnetic field energy or cosmic ray energy (Kamaya 1996).
It is possible that what is physically affected by magnetic fields
is gas density, and then the gas density determine SFE by the Schmidt
law, producing the observed $B$-SFE relation.

Anyway, the correlation between $B$ and SFE has actually been 
observed, and it is natural to consider this relation holds also 
in high redshift objects. [The chance probability to observe such
a correlation from an uncorrelated parent population is
only 12\% (Vall\'ee 1994).]
Therefore the following analysis is valid
unless some unknown processes violate this empirical
relation in high redshift objects.

\section{Magnetic Fields in Hierarchical Structure Formation}
As discussed above, it is suggested that spheroidal components
of galaxies formed by starbursts induced by strong magnetic fields.
In the following we consider the magnetic field generation and formation
of spheroidal systems in the framework of the standard cosmological 
structure formation in the CDM universe, and show that
spheroidal systems form only in relatively massive objects at
high redshifts. Recently it is discussed that galactic magnetic
fields are generated at the stage of collapse of protogalactic clouds
(Kulsrud et al. 1997). Although there are some other scenarios as
the origin of galactic magnetic fields, such as earlier field generation on 
cosmological scales or dynamo amplification after galaxies form
(see, e.g., Zweibel \& Heiles 1997), we assume here that magnetic
fields are generated during collapse of protogalactic clouds.
When a dark matter halo decouples from the expansion of the universe
and collapses into a virialized object, the gravitational energy of baryonic
gas is converted into turbulent motions and thermal energy by generated
shocks. It can be shown that a very weak magnetic field ($\sim
10^{-21}$ G), which has been generated by thermoelectric current before
collapse (i.e., the battery mechanism), 
is amplified up to a strength nearly in equipartition with
turbulent energy (Kulsrud \& Anderson 1992; Kulsrud et al. 1997). 
The time scale for equipartition is given by
$\sim r/\upsilon$, where $r$ is the length scale of the system and 
$\upsilon$ the turbulent velocity. (The essence of this result can be 
understood by a dimensional analysis.
The equation of magnetic field generation is $d{\bf B}/dt = {\rm rot}({\bf
v \times B})$, where ${\bf v}$ is the velocity field
of fluid, and when we estimate rot by the inverse of the system
scale, it is clear that the field evolution time scale is given by 
$\sim r/\upsilon$.) If we take the radius and three-dimensional 
velocity dispersion of a virialized halo as $r$ and $\upsilon$, 
it is straightforward to see that this time scale is given by 
the dynamical time scale of the halo. Then we can estimate 
the strength of magnetic fields by the turbulent energy density of 
baryonic gas in a collapsed object, which is determined by using the
well-known spherical collapse model (e.g., Peebles 1980). 
For a dark halo with
mass $M_h$ and formation redshift $z$, the turbulent energy density becomes
$\epsilon_B \sim (3 \Omega_B M_h \upsilon^2 / 8 \pi \Omega_0 r^3) =
6.4 \times 10^{-13} h^{8/3} M_{12}^{2/3} (1+z)^4 \Omega_B \rm \ erg
\ cm^{-3}$, where $h$ is the Hubble constant normalized at
100 km/s/Mpc, $M_{12} = M_h / (10^{12} M_\odot)$, and $\Omega_B$ and
$\Omega_0$ are the baryon and matter density in units of the critical density
in the universe. In the second expression above, we have assumed the 
Einstein-de Sitter universe ($\Omega_0 = 1$), but extension to other
cosmological models is easy.  According to the theory of
Kulsrud and Anderson (1992), we assume that one-sixth of the turbulent energy 
is converted into magnetic field energy and then the equipartition 
magnetic field becomes $B \sim 1.6 h^{4/3} M_{12}^{1/3} (1+z)^2
\Omega_B^{1/2} \ \mu$G. The observational properties of disk galaxies
including our Galaxy are well understood if they are considered
to have formed at $z \sim$ 0--1 (Mao \& Mo 1998), 
and hence this theory of magnetic
field generation gives a roughly correct strength for our Galaxy
with $M_h \sim 10^{12} M_\odot$. 

\section{Emergence of the Hubble Sequence}
Then spheroidal galaxies are expected to form at high redshifts
in massive dark matter halos, because 
magnetic field becomes stronger with increasing mass and
redshift as $B \propto M_h^{1/3} (1 + z)^2$. In order to discuss more
quantitatively, we equate the previously defined  
star formation efficiency (in \S 2) to $\nu
= (t_{SF})^{-1}$, where $t_{\rm SF}$ is the time scale on which interstellar
gas is converted into stars. The star formation rate $\dot M_*$ in a galaxy 
is given by $\dot M_* = M_{\rm gas}/t_{\rm SF}$, where $M_{\rm gas}$
is the mass of interstellar gas in the galaxy. As mentioned earlier,
the observed relation between $B$ and $\nu$ is $\nu \propto
B^q$ with $q \sim 7.7 \pm 2.6$.
We set the normalization of this relation by SFE at our Galaxy.
The star formation rate in our Galaxy is about a few $M_\odot$/yr
and mass of interstellar gas in the disk is about 
$6 \times 10^9 M_\odot$ at present, suggesting that $t_{\rm SF} \sim 2$ Gyr
at $B \sim 3 \mu$G (e.g., Binney \& Tremaine 1987). 
We also normalize the strength of $B$ to 3 $\mu$G
for a halo whose baryonic mass is $6 \times 10^{10} M_\odot$ and
whose virialization occurred at $z=1$, supposing our Galaxy. 
Then we can calculate the 
ratio of baryonic dynamical time to $t_{\rm SF}$, $\Gamma \equiv 
t_{\rm dyn} /
t_{\rm SF}$, which can be considered as a criterion for spheroidal formation. 
In calculation of $t_{\rm dyn}$, we use a baryon density
$\lambda^{-3}$ times higher than the value at virialization, considering
the contraction of baryons due to cooling and dissipation, 
where $\lambda \sim 0.05$
is typical dimensionless angular momentum of dark haloes 
(Warren et al. 1992).  
If $\Gamma \gg 1$, spheroidal systems are expected to form.

\begin{figure}
  \begin{center}
    \leavevmode\psfig{file=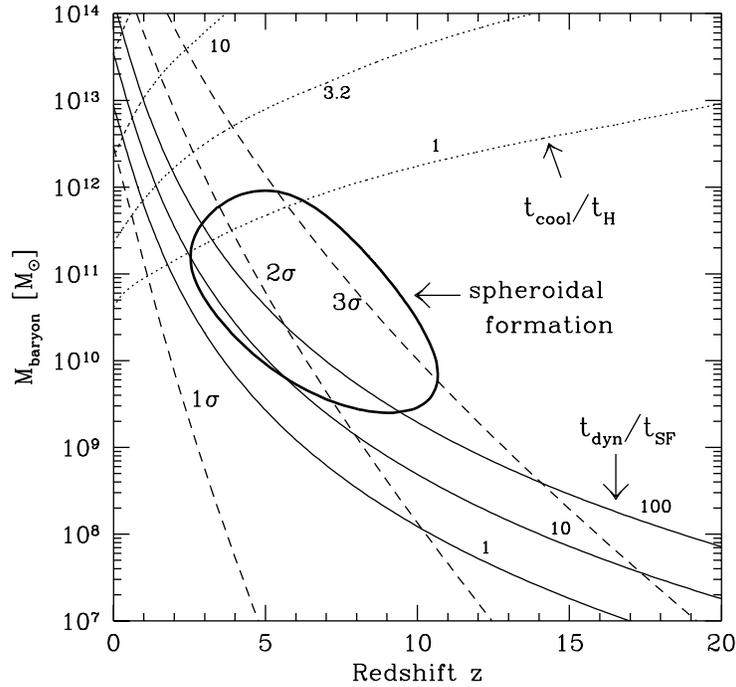,width=10cm}
  \end{center}
\caption{
Thin solid lines show contours of the ratio of dynamical time
scale to star formation time scale, $t_{\rm dyn}/t_{\rm SF}$
= 1, 10, and 100, as functions of baryon mass and formation redshift
of cosmological objects. When $t_{\rm dyn}/t_{\rm SF} \gg 1$, spheroidal
systems are expected to form. Dashed lines show the mass-redshift
relation for objects with a given $n$-$\sigma$ fluctuation
for $n$ = 1, 2, and 3, where $\sigma$ is the root-mean-square of
the density fluctuations in the universe at a given mass scale.
The value of $n$ represents statistical abundance of collapsed objects.
Dotted lines are contours of the ratio of
cooling time to the Hubble time at given redshifts, for
$t_{\rm cool}/t_H$ = 1, 3.2, 10. Galaxy formation is inhibited for
objects with $t_{\rm cool}/t_H \protect\gtilde 1$. The region in which 
spheroidal galaxies form is schematically shown by the 
thick solid line. A $\Lambda$-dominated flat
universe with cosmological parameters of 
$(h, \Omega_0, \Omega_\Lambda, \sigma_8) = (0.7, 0.3, 0.7, 1)$ is
assumed, where $\Lambda$ is the cosmological constant.
}
\label{fig:hubble}
\end{figure}

\begin{figure}
  \begin{center}
    \leavevmode\psfig{file=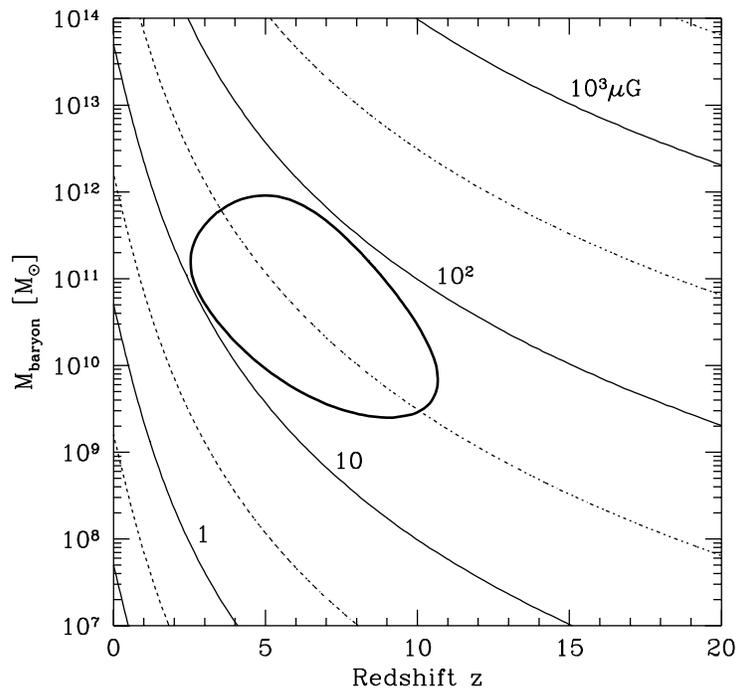,width=10cm}
  \end{center}
\caption{The same as Fig. \protect\ref{fig:hubble}, but
for the contour of magnetic field strengths in gravitationally bound
objects. The indicated values associated to thin solid contours
are magnetic field strengths in $\mu G$, and dotted lines are
contours with a step of 0.5 in $\log B$. The magnetic field strength
is normalized to 3 $\mu$G at $M_{\rm baryon} = 6 \times 10^{10} M_\odot$
and $z = 1$ supposing our Galaxy, and this normalization is very close
to the equipartition field with the turbulent energy density of
protogalactic clouds (see text). The region in which 
spheroidal galaxies form is schematically shown by the 
thick solid line (see Fig. \protect\ref{fig:hubble}). 
}
\label{fig:mag}
\end{figure}

In Figure \ref{fig:hubble}, 
we have plotted the contour of $\Gamma$ = 1, 10, and 100
by thin solid lines
in a plane of baryon mass of a collapsed object ($M_{\rm baryon}$)
and its formation redshift, assuming $q=5$.
(Observed mass of galaxies is considered to be mainly baryonic mass.)
For reference, we have also plotted the contour of magnetic field
strengths in Fig. \ref{fig:mag}.
We have used a flat, $\Lambda$-dominated cosmological model,
where $\Lambda$ is the cosmological constant, with 
the standard cosmological parameters of 
$(h, \Omega_0, \Omega_\Lambda, \sigma_8) = (0.7, 0.3, 0.7, 1)$.
This universe is well consistent with various observations such as
ages of globular clusters, high redshift type Ia supernovae
(Perlmutter et al. 1998), 
cosmic microwave background (Bunn \& White 1997), and abundance of clusters of
galaxies (Kitayama \& Suto 1997). In the following we discuss the
emergence of the Hubble sequence using Fig. \ref{fig:hubble}.
Objects lying above the thin solid lines are expected to form spheroidals, but
we have to check how such objects are typical in the context of
cosmological structure formation in the CDM universe. For this purpose we
have plotted mass scale of $n$-$\sigma$ fluctuation as defined in 
Blumenthal et al. (1984),
by dashed lines. The lines are defined by $n \sigma(M_h, z) = \delta_c$, where 
$\sigma(M_h, z)$ is the root-mean-square of density fluctuation predicted by
linear theory in the CDM universe 
at scale of $M_h$ and at redshift $z$ (Peacock \& Dodds 1994; 
Sugiyama 1995), and $\delta_c \sim 1.69$ is a 
critical density contrast at which an object virializes (Peebles 1980).
The three dashed lines correspond to $n$ = 1, 2, and 3, and these lines 
represent typical mass scales of collapsed objects as a function of $z$.
(For larger $n$, the cosmological abundance of objects is statistically 
suppressed as $\propto e^{-n^2/2}$.)
In the region where the dashed lines are above the solid lines, 
spheroidal galaxies can form as cosmologically typical objects. 
Figure \ref{fig:hubble} shows that objects with $M_{\rm baryon}
\gtilde 10^{9-10} M_\odot$ with $\sim 2$--$3 \sigma$ fluctuation will form
spheroidal systems because $\Gamma \gg 1$, at high redshifts 
of $z \sim$ 3--10 with $B \sim 50 \mu$G (see also Fig. \ref{fig:mag}). 
It is interesting to note that
this magnetic field strength is roughly the same with that in the
starbursting nucleus of M82, as mentioned in \S 2.
Spheroidal formation with $M_{\rm baryon} \ltilde
10^{9-10} M_\odot$ is inhibited because of low $\Gamma$ for
cosmologically typical objects. On the other hand, if $M_{\rm baryon}$
is larger than $\sim 10^{12} M_\odot$, formation of galaxies is 
inhibited by too long cooling time ($t_{\rm cool}$) 
compared to the Hubble time at each redshift ($t_H$),
as shown by dotted lines which are contours of
$t_{\rm cool}/t_H$ (Blumenthal et al. 1984). Therefore, our model can explain 
the observed mass range ($10^{10-12} M_\odot$)
and old stellar populations seen in spheroidal
galaxies. The region for spheroidal galaxy formation is schematically
indicated in Figs. \ref{fig:hubble} and \ref{fig:mag} by the thick solid line.
After spheroidal galaxy formation, gas accretion onto some
of them is possible at lower redshifts. Since such gas accreting 
recently is located below the thin solid lines in Figure 
\ref{fig:hubble}, it results in
disk formation of spiral galaxies at $z \ltilde 1$. On the other hand,
if recent gas accretion is negligible, such object will be seen as
present-day elliptical galaxies.

Since solid lines (constant $\Gamma$) and 
the dashed lines ($n$-$\sigma$ lines) are approximately parallel
in the region of spheroidal formation,
our scenario for the origin of the Hubble sequence predicts that
elliptical galaxies should lie on a constant $n$-$\sigma$ density
fluctuation with $n \sim$ 2--3, and with decreasing $n$, the type
of galaxies becomes later in the Hubble sequence, i.e., 
from early to late spiral, and then into irregular galaxies
at $\sim 1 \sigma$ fluctuation. In fact, this trend is exactly
what has been observed in galaxies (Blumenthal et al. 1984;
Burstein et al. 1997), giving a further support for
our scenario. Most properties or relations observed in present-day
galaxies, such as the Tully-Fisher or Faber-Jackson relation, 
existence of the fundamental plane for galaxies and their distribution
on it, can be explained by formation of early type
galaxies from higher $n$-$\sigma$ fluctuations than later Hubble types
(Burstein et al. 1997).
The density-morphology relation, which is the correlation between
galaxy types and number density, is also explained; higher $n$-$\sigma$
fluctuations occur preferentially in denser regions destined to become
rich clusters, and hence one expects to find more ellipticals there,
as is observed (Blumenthal et al. 1984). 
Higher $n$-$\sigma$ objects are expected to show stronger
spatial clustering than lower ones (e.g., Mo \& White 1996), 
and it is consistent with the stronger
clustering observed for elliptical galaxies than late type galaxies
(Loveday et al. 1995).
Various observations suggest that giant galaxies formed at higher
redshifts of $z \gtilde 3$, and were then followed by a sequence of less 
and less massive galaxies forming at lower and lower redshifts, leading
down to the formation of dwarfs at recent ($z \ltilde 0.5$)
(Fukugita, Hogan \& Peebles 1996; Sawicki, Lin, \& Yee 1997). 
The proposed scenario gives an explanation for this trend which is sometimes
termed as ``downsizing'', otherwise it seems opposite
to the expectation in the CDM universe.

\section{Discussion \& Conclusions}
For very low mass objects, star formation is strongly suppressed
by the absence of physical triggers when
magnetic fields are weak, and this may give an explanation for the fact
that the faint end of luminosity function of galaxies is much flatter
than  expected from the mass function of dark haloes
(Kauffmann et al. 1993; Baugh et al. 1996). Because of the strong 
dependence of star formation activity on magnetic fields, 
stellar luminosity of galaxies would quite rapidly decrease with
decreasing mass of galaxies. Then 
it is expected that, in the faint end of galaxy luminosity function,
the mass of galaxies does not so change compared to the change in
luminosity. Since the number density of objects is determined by
mass of objects, the faint end is expected to be relatively flat,
as observed. Irregular galaxies or late-type galaxies are expected
to show a wide range of star formation activity within a narrow
range of galaxy masses depending on
the magnetic field strength in them, and in fact such a trend has
been observed (Hunter 1997).

It is well known that galaxy interactions or mergers induce intensive
starbursts, but the mechanism which triggers starbursts in galaxy 
interactions is still poorly known. Mergers are followed by cloud
collisions leading to high gas density, strong turbulences and 
strong magnetic fields, and ending in gas collapse and star formation
in clouds. Therefore the hypothesis presented in this letter, i.e.,
strong dependence of star formation activity on magnetic field strengths,
may also be important in the merger-induced starbursts.
Some fraction of elliptical galaxies may have formed by such a process,
but we have argued that the observed trends of ellipticals, i.e.,
being massive and old, originate mainly from properties of
gravitationally bound objects in the standard theory of cosmological structure
formation.

We have presented a new idea for the origin of the Hubble sequence of 
galaxies. The key of this scenario is the strong dependence of
star formation activity on average magnetic fields in galaxies,
which is just a speculation from a theoretical point of view, but motivated
by several observational facts about magnetic fields in nearby galaxies.
This only one speculation provides a simple
explanation for the surprisingly narrow dispersion in the
magnetic field strengths observed in spiral galaxies, and also 
for most properties of galaxies seen along the Hubble sequence. To verify this
``magnetic galaxy formation'' 
scenario as the origin of the Hubble sequence, it is
indispensable to confirm observationally that starbursts are 
actually triggered by stronger magnetic fields. Comprehensive measurements of
galactic magnetic fields in larger samples of galaxies are necessary,
especially for starburst galaxies at local as well as at high redshifts.

The author would like to thank T. Kitayama, M. Shimizu, and an anonymous
referee for useful comments and discussions.

\end{document}